\documentstyle[12pt,aasms4]{article}

\begin{document}
\title{Non-Gaussianity and the recovery of the mass power spectrum
from the Ly$\alpha$ forest}

\author{Long-long Feng\footnote{fengll@physics.arizona.edu}}

\affil{Center for Astrophysics, University of Science and
Technology of China, Hefei, Anhui 230026, P.R. China}

\and

\author{Li-Zhi Fang\footnote{fanglz@physics.arizona.edu}}

\affil{Department of Physics, University of Arizona, Tucson, AZ
85721}

\begin{abstract}

We investigate the effect of non-Gaussianity on the reconstruction
of the initial mass field from the Ly$\alpha$ forest. We show that
the transmitted flux of QSO absorption spectra are highly
non-Gaussian in terms of the statistics, the kurtosis spectrum and
scale-scale correlation. These non-Gaussianities can not be completely
removed by the conventional algorithm of Gaussianization, and
the scale-scale correlations are largely retained in the
mass field recovered by the Gaussian mapping. Therefore, the mass
power spectrum recovered by the conventional algorithm is
systematically lower than the initial mass spectrum on scales
at which the local scale-scale correlation is substantial. To
reduce the non-Gaussian contamination, we present two methods. The
first is to perform the Gaussianization scale-by-scale using the
discrete wavelet transform (DWT) decomposition. We show that the
non-Gaussian features of the Ly$\alpha$ forest basically will no
longer exist in the scale-by-scale Gaussianized mass field. The
second method is to choose a proper orthonormal basis
(representation) to suppress the effect of the non-Gaussian
correlations. In the quasilinear regime of cosmic structure
formation, the DWT power spectrum is efficient for suppressing the
non-Gaussian contamination. These two methods significantly improve the
recovery of the mass power spectrum from the Ly$\alpha$ forest.

\end{abstract}



\section{Introduction}

Since  high resolution QSO spectra became available, the
transmitted flux in QSO spectra, or the Ly$\alpha$ forest,
offers an unprecedented opportunity to study the large-scale
structure of the universe and its evolution at redshifts beyond
the galaxy redshift catalog (e.g. Bi \& Davidsen 1997 and
references within). A basic goal of this study is to reconstruct
the initial mass field. Assuming that these objects trace the
underlying matter field in some way, it seems to be possible to
trace the evolution of mass field back in time. Because the
initial mass field is expected to be Gaussian in many models of
the origin of fluctuations, reconstructing the initial mass
fluctuations is synonymous with recovering the mass power spectrum
(e.g. Croft et al. 1999.)

The recovery of power spectrum from the Ly$\alpha$ forests relies
upon two theoretical conjectures. The first is to assume that the
transmitted flux of a QSO Ly$\alpha$ absorption spectrum  is a
{\it point-to-point} tracer of the underlying dark matter
distribution. The Ly$\alpha$ forest has been successfully modeled
by the absorption of the ionized intergalactic gas, of which the
distribution is continuous, and locally determined by the
underlying dark matter distribution (Bi 1993; Fang et al. 1993;
Bi, Ge \& Fang 1995; Hernquist et al 1996; Bi \& Davidsen 1997;
Hui, Gnedin \& Zhang 1997). Thus, the transmitted flux of a QSO
absorption spectrum at a given redshift depends only on the mass
density of dark matter at the position corresponding to the
redshift.

The second assumption is that the initial mass field can be
recovered from the flux of QSO spectrum by the Gaussianization
algorithm (Weinberg 1992; Croft et al. 1998). With this method,
the shape of 1-D initial Gaussian density field with an arbitrary
normalization can be recovered approximately from the observed
flux by a point-to-point Gaussian mapping if the relation between
flux and mass density is monotonous, i.e. the higher the
underlying mass density, the stronger the Ly$\alpha$ absorption.
The monotony would be a good approximation in the weak nonlinear
or quasilinear evolutionary regime of the cosmic clustering.

This paper is trying to study the influence of the non-Gaussianity
of the Ly$\alpha$ forest on the recovery of the initial power
spectrum. It is motivated by the recently systematic detection of
non-Gaussianity of the Ly$\alpha$ forests. Despite it is well
known that the two-point correlation function of the Ly$\alpha$
absorption lines is quite weak, the distribution of these lines
does show non-Gaussian behavior. For instance, it has been
well known about ten years ago that the distribution of the
nearest neighbor Ly$\alpha$ line intervals is different from a
Poisson process (Duncan, Ostriker, \& Bajtlik 1989; Liu and Jones
1990; Fang, 1991). Recently, the detection of the spectrum of
higher order cumulants (Pando \& Fang 1998a) and the scale-scale
correlations (Pando et al. 1998) of the Ly$\alpha$ forests implies
 systematic non-Gaussianity on scales as large as about 10
h$^{-1}$ Mpc. The abundance of the Ly$\alpha$ line ``clusters"
identified with respect to the richness is also found to be
significantly different from a Gaussian process (Pando \& Fang
1996.)

According to the philosophy of the Gaussianization reconstruction,
all the non-Gaussian features of the Ly$\alpha$ forests are not
initial. It should be removed by the Gaussianization of the flux
of QSO spectra. The recovered mass field should be Gaussian.
The algorithm of Gaussian mapping is designed for removing the
non-Gaussianities of the flux, and recovering a Gaussian mass
field.

The idea of Gaussianization is exquisite. However, we will show
that even though the current algorithm of Gaussian mapping does map
the distribution of the flux value into a Gaussian probability
distribution function (PDF), the above-mentioned non-Gaussianities
of the Ly$\alpha$ forests still remain largely in the Gaussianized
flux. Namely, the recovered field is not linear and Gaussian, but
contaminated by the non-Gaussian behavior in the Ly$\alpha$
forest.

It has been recognized that the estimation of power spectrum is
significantly affected by the non-Gaussian behavior, such as the
correlation between band-averaged power spectra which is
essentially the scale-scale correlation (e.g. Meiksin \& White
1999). Therefore, with the current algorithm of Gaussianization,
the recovered power spectrum is distorted by the non-Gaussianity
of the Ly$\alpha$ forests. The question is then raised: how to
improve the algorithm of Gaussianization in order to recover a
mass field exempted from the non-Gaussianity of the Ly$\alpha$
forest? Or how to suppress the effect of the non-Gaussian
contamination in estimation of the power spectrum? We try to
investigate these problems in this paper.

This paper is organized as follows. In \S 2, using popular cold
dark matter models, we present the non-Gaussian features in the
transmitted flux of the Ly$\alpha$ forests. In \S 3, we
demonstrate that the non-Gaussianity of mass field recovered by
the conventional Gaussianization algorithm is about the same order
as the original non-Gaussianity. Two alternatives which yield
better Gaussianization are then proposed. In \S 4, the distortion
of power spectrum by the non-Gaussianity is shown, and a possible
way of suppressing the non-Gaussian effect on the power spectrum
detection, i.e. properly choosing representation of the power
spectrum, is suggested. We conclude this paper with a discussion
of our findings in \S 5.

\section{The Non-Gaussian features of the Ly$\alpha$ forests}

\subsection{Samples of the Ly$\alpha$ forests}

To investigate the recovery method of mass power spectrum, we
generate the simulation samples of the Ly$\alpha$ forest in the
semi-analytic model of the intergalactic medium (IGM) developed by
H.G. Bi et al. (Bi 1993; Fang et al. 1993; Bi, Ge \& Fang 1995; Bi
\& Davidsen 1997.) This model can approximately fit most observed
features of the Ly$\alpha$ forest, including the column density
distribution and the number density of the Ly$\alpha$ forest
lines; the distribution of the equivalent widths and their
redshift dependence; the clustering and the Gunn-Peterson effect.
Moreover, in this model, the relations among the dark matter
field, the flux of the Ly$\alpha$ absorption and the power
spectrum of reconstructed initial mass fields are under control.
It would be very useful to reveal the problems of the
reconstruction.

The model was described in detail in the above listed references.
We now give a brief account of, especially, the fundamental
physics underlying in this model. The basic assumption of the
model is that the density distribution of the baryonic diffuse
matter $n({\bf x})$ in the universe is determined by the
underlying dark matter density distribution via a lognormal
relation as
\begin{equation}
n({\bf x}) = n_0 \exp \left [ \delta _0 ({\bf x}) -
  \frac{<\delta_0 ^2>}{2} \right ],
\end{equation}
where $n_0$ is the mean number density, and $\delta _0({\bf x})$
is a Gaussian random field derived from the density contrast
$\delta _{DM}$ of the dark matter by:
\begin{equation}
\delta _0({\bf x}) = \frac{1}{4\pi x_b ^2} \int \frac{\delta _{DM}
({\bf x}_1)} {|{\bf x} - {\bf x}_1|}e^{-\frac{|{\bf x} - {\bf
x}_1|}{x_b}} d{\bf x}_1
\end{equation}
in the comoving space, or
\begin{equation}
\delta_0 ({\bf k}) = \frac{\delta _{DM} ({\bf k})}{1+x_b^2k^2}
\end{equation}
in the Fourier space. To take into account the effect of redshift distortion,
the peculiar velocity field along the line of sight is also calculated by the
simulation model (Bi 1993; Fang et al. 1993; Bi\& Davidsen 1997.)

The Gaussian field $\delta _{DM}$ is produced in a cold dark
matter model. To account for the baryonic effect on the transfer
function, we adopt the fitting formula for power spectrum
presented by Eisenstein \& Hu (1999). Because the goal of this
paper is mainly on examining the recovery method of power
spectrum, we will not take into account the variants of CDM
family, but only the ``standard" one, i.e. the flat model
($\Omega_0=1.0$) normalized by the 4-year COBE data, and the
$\Gamma=\Omega_0h$ is taken to be 0.3, where $h$ denotes the
normalized Hubble parameter, and $\Omega_0$ is the cosmological
density parameter of total mass.  This model is compatible with
the galaxy correlation observed on scales of $\sim 10$ h$^{-1}$
Mpc (Efstathiou et.al. 1992). The baryonic fraction in the total
mass was fixed by the constraint from the primordial
nucleosynthesis of $\Omega_b=0.0125$  h$^{-2}$ (Walker et.al,
1991).

The factor $x_b$ in Eq. (2) is the Jeans length of IGM given by
\begin{equation}
x_b \equiv \frac{1}{2\pi H_0} \left [\frac{2\gamma k T_m } {3\mu
m_p \Omega (1+z)} \right ] ^{\frac{1}{2}},
\end{equation}
where $T_m$ and $\mu$ are the density-average temperature and
molecular weight of the IGM respectively, and $\gamma$ is the
ratio of specific heats. The thermal equation of state of IGM is
assumed to be polytropic, $T\propto n^{\gamma-1}$ with
$\gamma=4/3$.

The lognormal relation, Eq.(1), has the following property: (1)
When fluctuations are small, i.e. $( n/n_0 -1) \simeq \delta _0$,
Eq. (1) is just the expected linear evolution of the IGM; (2) On
small scales as $|{\bf x} - {\bf x}_1| \ll x_b$, Eq. (1) becomes
the well-known isothermal hydrostatic solution, which describes
highly clumped structures such as intracluster gas, $n \propto
\exp (-\mu m_p \psi _{DM}/\gamma k T)$, where $\psi _{DM}$ is the
dark matter potential (Sarazin \& Bahcall 1977).

The absorption optical depth at observed wavelength $\lambda$ is
\begin{equation}
\tau (\lambda)  =  \int ^{t_0} _{t_{qso}}
  \sigma \left (\frac{c}{\lambda_{\alpha}}\frac{1+z}{1+z_0}\right )
   n_{HI} (t) dt,
\end{equation}
where $z_0=(\lambda/\lambda_{\alpha})-1$, $t_0$ denotes the
present time, $t_{qso}$ is the time corresponding to the redshift
$z_{qso}$ of the QSO, and so does for the relation between $t$ and
$z$; $\sigma$ is the absorption cross section at the Ly$\alpha$
transition, and $\lambda_{\alpha}=1216 \AA$ represents the
Ly$\alpha$ wavelength. The density of the neutral hydrogen atoms,
$n_{HI}$, can be found from $n$ by the cosmic abundance of
hydrogen, and photoionization equilibrium (Bi, Ge \& Fang 1995.)

Obviously in this model, the relation between the transmitted
flux, $F(\lambda)=e^{-\tau}$, and $n({\bf x})$ or $\delta_0({\bf
x})$ is basically local. The non-locality is only caused by the
width of the absorption cross section $\sigma$, and the peculiar
velocity of the neutral hydrogen. Therefore, $F$ is approximately
a point-to-point tracer of mass fluctuation $\delta_0({\bf x})$.
Moreover, the $F$ is monotonically related to $\delta_0$, and
then, to the density contrast $\delta _{DM}$ on scales larger than
$x_b$.

In this paper, we produce the simulation samples in the redshift
range $z=2.066\sim 2.436$ with $2^{14}$ pixels. The corresponding
simulation size in the CDM model is 189.84 h$^{-1}$Mpc in comoving
space which is long enough to incorporate most of the fluctuation
power. The selection of this redshift range is to compare the
simulation with the Keck spectrum of HS1700+64. The spectrum of
HS1700+64 ranges from 3723.012\AA to 5523.554\AA with the
resolution $3$ kms$^{-1}$, or totally 55882 pixels in which the
first $2^{14}$ pixels are chosen here. These data have been used
for testing the model considered in Bi \& Davidsen (1997).

\subsection{The skewness and kurtosis spectra of the transmitted flux}

If appropriate parameters of the intergalactic UV background are
adopted, the lognormal IGM model described in \S 2.1 could explain
successfully many observed properties of the Ly$\alpha$ forest and
their evolution from redshift 2 to 4. Now we show that it also
works against tests of the non-Gaussian features.

We use the wavelet transform to analyze the non-Gaussian behavior
of the transmitted flux $F$.  As a 1-D field, the flux
$F(\lambda)$ in the wavelength range of
$L=\lambda_{max}-\lambda_{min}$ is subject to a discrete wavelet
transform (DWT) as
\begin{equation}
F = \bar{F} + \sum_{j=0}^{\infty} \sum_{l= 0}^{2^j -1}
  \tilde{\epsilon}_{j,l} \psi_{j,l}(\lambda)
\end{equation}
where $\psi_{j,l}(x)$, $j=0,1,...$, $l=0...2^j-1$, is an
orthogonal and complete set of the DWT basis (the details of the
DWT, see e.g. Fang \& Thews 1998). The wavelet basis
$\psi_{j,l}(x)$ is localized both in the physical space and the
Fourier (scale) space. The function $\psi_{j,l}(x)$ is centered at
position $lL/2^j$ of the physical space, and at wavenumber $2\pi
\times 2^j/L$ of the Fourier space. Therefore, the wavelet
function coefficients (WFCs), $\tilde{\epsilon}_{j,l}$, have two
subscripts $j$ and $l$. They describe the fluctuation of the flux
on scale $L/2^j$ at position $lL/2^j$. To be more specific, we
will use the Daubechies 4 wavelet in this paper, although all
conclusions are not affected by this particular choice as long as
a compactly supported wavelet basis is used.

The WFC, $\tilde{\epsilon}_{j,l}$, is computed by the inner
product of
\begin{equation}
\tilde{\epsilon}_{j,l}= <F * \psi_{j,l}>.
\end{equation}
Since the DWT bases are complete, the WFCs contain all information
of the flux.

Note that the $\psi_{j,l}(x)$ are orthogonal with respect to the
position index $l$, and therefore, for an ergodic field, the $2^j$
WFCs at a given $j$, i.e. $\tilde{\epsilon}_{j,l}$,
$l=0,1...2^j-1$, can be treated as independent measures of the
flux field. The $2^j$ WFCs, $\tilde{\epsilon}_{j,l}$, from {\it
one} realization of $F(\lambda)$ can be employed as a statistical
ensemble. In other words, when the fair sample hypothesis holds
(Peebles 1980), an ensemble average can be estimated equivalently by
averaging over $l$, i.e. $\langle \tilde{\epsilon}_{j,l} \rangle
\simeq (1/2^j)\sum_{l=0}^{2^j-1}\tilde{\epsilon}_{j,l}$, where
$\langle ...\rangle$ denotes the ensemble average. The
distribution of $\tilde{\epsilon}_{j,l}$ represents approximately
the one-point distribution of the WFCs at a given scale $j$.

The non-Gaussianity of the flux $F(\lambda)$ can be directly
measured by the deviation of the one-point distribution from a
Gaussian distribution. For this purpose, we calculate the cumulant
moments defined by
\begin{equation}
I^2_j= M^2_j,
\end{equation}
\begin{equation}
I^3_j = M^3_j,
\end{equation}
\begin{equation}
I^4_j= M^4_j - 3 M^2_j M^2_j,
\end{equation}
\begin{equation}
I^5_j= M^5_j - 10 M^3_jM^2_j,
\end{equation}
where
\begin{equation}
M^n_j \equiv \frac{1}{2^j} \sum_{l=0}^{2^j-1}
(\tilde{\epsilon}_{j,l} - \overline{\tilde{\epsilon}_{j,l}})^n.
\end{equation}

The second order cumulant moment gives the DWT power spectrum (\S
4) (Pando \& Fang 1998). For Gaussian fields all the cumulant
moments higher than order 2 are zero. Thus one can measure the
non-Gaussianity by $I^n_j$ with $n>2$. We call $I^n_j$ the DWT
spectrum of $n$-th cumulant. The cumulant measures $I^3_j$ and
$I^4_j$ are related to the well known skewness and kurtosis,
respectively, defined by
\begin{equation}
S_j \equiv \frac{1}{(I_j^2)^{3/2}} I^3_j,
\end{equation}
\begin{equation}
K_j  \equiv   \frac{1}{(I_j^2)^2} I^4_j.
\end{equation}

Using the skewness and kurtosis spectra as statistical indicators,
a significant non-Gaussian behavior has been found in the distribution
of Ly$\alpha$ forest lines (Pando \& Fang 1998a). The skewness
and kurtosis spectra of the transmitted flux in 100 simulated
samples are shown in Figs. 1 and 2, respectively. To assess the
statistical significance, the 95\% confidence range from 100
realizations of Gaussian noise are also displayed in these figures.
Clearly, the kurtosis spectrum of simulated $F$ shows difference
from the Gaussian noise spectra on the scales $j \geq 8$
(or $\leq 1.5$ h$^{-1}$ Mpc) with 95\% confidence. The skewness
spectrum does not show significant difference from the Gaussian
noise till $j=11$ ($\sim 100$ h$^{-1}$ kpc). These results are
qualitatively consistent with that for the observed forest line
distributions. In addition, the skewness and kurtosis spectra of
the flux of HS1700+64 are also presented in Figs. 1 and 2. Obviously,
the CDM model is in excellent agreement with the observation.

\subsection{The scale-scale correlations of the transmitted flux}

The scale-scale correlations measure the correlations between the
fluctuations on different scales (Pando et al. 1998, Pando,
Valls--Gabaud, \& Fang 1998, Feng, Deng \& Fang 2000). This
non-Gaussianity is independent from the higher order cumulants (\S
2.2), which are only of $j$ dependent.  A simplest measure of the
scale-scale correlation is given by
\begin{equation}
C^{p,p}_j = \frac{2^{(j+1)}\sum_{l=0}^{2^{j+1}-1}
\tilde{\epsilon}^p_{j;[l/2]} \; \tilde{\epsilon}^p_{j+1;l}} {\sum
\tilde{\epsilon}^p_{j,[l/2]} \sum \tilde{\epsilon}^p_{j+1;l}}
\end{equation}
where $p$ is an even integer, and [\hspace{2mm}]'s denote the
integer part of the quantity. Because $Ll/2^j = L2l/2^{j+1}$, the
position $l$ at scale $j$ is the same as the positions $2l$ and
$2l+1$ at scale $j+1$. Therefore, $C_j^{p,p}$ measures the
correlation between fluctuations on scale $j$ and $j+1$ at the
{\it same} physical point. For Gaussian fields, $C_j^{p,p}=1$.
$C_j^{p,p}>1$ corresponds to the positive scale-scale correlation,
and $C_j^{p,p}<1$ to the negative case. One variant of the above
definition is
\begin{equation}
C^{p,p}_{j, \Delta l} = \frac{2^{(j+1)}\sum_{l=0}^{2^{j+1}-1}
\tilde{\epsilon}^p_{j;[l/2]+\Delta l} \;
\tilde{\epsilon}^p_{j+1;l}} {\sum \tilde{\epsilon}^p_{j,[l/2]}
\sum \tilde{\epsilon}^p_{j+1;l}}.
\end{equation}
This statistics is for measuring the correlations between
fluctuations on scales $j$ and $j+1$, but at different positions,
i.e. the fluctuation at scale $j$ is displaced from the $j+1$
fluctuation by a distance $\Delta l L/2^j$.

The scale-scale correlation $C_j^{2,2}$ calculated from the
simulated transmitted flux and HS1700+64 are shown in Fig. 3.
Clearly, the values of $C_j^{2,2}$ are significantly larger than
unity and well above the Gaussian noise spectra on all the scales
$j \geq 7$. This result is also qualitatively in agreement with
the scale-scale correlation of the Ly$\alpha$ forests (Pando et
al. 1998.). Figure 3 also indicates that the model of \S 2.1 is
still in a good shape of fitting the observed non-Gaussian
correlation.

Similar to Eq.(15) for the correlation between scales $|j-j'| =1$,
one may define, in principle, the correlation between two arbitrary
scales with $|j-j'| >1$. However, for hierarchical clustering the
scale-scale correlation is quantified mainly by $|j-j'| =1$.
Therefore, we will not calculate the scale-scale correlations for
$|j-j'| >1$.

\section{The Non-Gaussian features of the Gaussian-recovered mass fields}

\subsection{Non-Gaussianity after Gaussianization}

The cosmological reconstruction is to extract the power spectrum
of the initial linear mass fluctuations from the observed
distribution of various tracers of the evolved density field.  The
algorithm of Gaussianization was designed for recovering the
primordial density fluctuations from an observed galaxy
distribution (Weinberg 1992). This method has been recently
applied to recovering the linear density field and its power
spectrum from the observed transmitted flux $F$ of QSO
absorption spectra (Croft et al 1998, 1999).

The key step of the Gaussianization algorithm is a pixel-to-pixel
mapping from an observed flux $F$ into the density contrast
$\delta$. The probability distribution function (PDF) of the
observed transmitted flux $F$ is generally non-Gaussian, while the
PDF of the initial density contrasts $\delta \equiv (n/n_0)-1$ is
assumed to be Gaussian in large variety of galaxy formation
models. The relation between $F=\exp(-\tau)$ and $\delta$ is
monotonic, i.e. high initial density $\delta$ pixels evolved into
high $\tau$ pixels, low initial density pixels into low $\tau$
pixels. Thus, using the observed $F$, one can sort out the total N
pixels by the amount of $F$ in the ascending order: the pixel with
lowest $F$ is labeled by 1st, the next higher $F$ pixel is labeled
by 2nd, and so on. For the n-th pixel, we then assign the density
contrast $\delta$, which is given by the solution of the equation
$(2\pi)^{-1/2}\int_{-\infty}^{\delta}\exp(-x^2/2)dx= n/N$. Thus,
the Gaussian mapping produces a mass field with the same rank
order as the flux but with a Gaussian PDF of $\delta({\bf x})$.
The overall amplitude of the recovered power spectrum should be
determined by a separate procedure.  For instance, we may set up
the initial condition by using the recovered spectrum, evolve the
simulation to the observed redshift and then normalize the
spectrum by requiring that the simulation reproduces the observed
power spectrum of the transmitted QSO flux. This amplitude
normalization is model-dependent.

We apply the Gaussianization to 100 simulation samples of the QSO
transmitted flux, and measure the skewness and kurtosis spectrum
as well as the scale-scale correlation. The results are displayed
in Figs. 4 - 6. For comparison, the non-Gaussian spectra of the
flux in Figs. 1 - 3. are also plotted correspondingly. Figs. 4 - 6
show that the Gaussianized flux still largely exhibits non-Gaussian
features. Especially, the scale-scale correlations of the
Gaussianized field is as strong as the pre-Gaussianized flux on
scales $j \geq 10$. That is, the recovered density field is
seriously contaminated by the non-Gaussianities in the original
flux.

\subsection{The efficiency of the conventional Gaussianization}

The reason for the lower efficiency of the conventional Gaussian
mapping (\S 3.1) is simple. The initial Gaussian random mass field
is assumed to be a superposition of independent modes, of which
the PDFs are Gaussian. For instance, in the Fourier
representation, all Fourier modes of a Gaussian mass field are
Gaussian, i.e. they have Gaussian PDF of the amplitudes and
randomized phases. The conventional algorithm considered only the
Gaussianization of one variable, $\delta$. It does not guarantee
the Gaussianization of the amplitudes and phases of all relevant
modes. In other words, the Gaussian mapping algorithm will work
perfectly for a system with one stochastic variable, but not so for
a field.

Alternatively, this problem can also be seen via the DWT
representation. Using eq.(1), any 1-D mass field given by density
contrast $\delta(x)$ ($\bar{\delta}=0$) can be decomposed with
respect to a DWT basis as
\begin{equation}
\delta(x) = \sum_{j=0}^{\infty} \sum_{l= 0}^{2^j -1}
  \tilde{\epsilon}^M_{j,l} \psi_{j,l}(x),
\end{equation}
where the superscript $M$ means mass. Equation (17) represents a
linear superposition of modes $\psi_{j,l}$. As has been pointed
out in \S2.2, for a given $j$, the 2$^j$ WFCs
$\tilde{\epsilon}^M_{j,l}$ form a statistical ensemble. The
distribution of the 2$^{j}$ WFCs gives the one-point distribution
of the amplitude of mode at the scale $j$. For the initial
Gaussian mass field, these one-point distributions should be
Gaussian. Obviously, the Gaussian PDF of $\delta$ does not imply
that the one-point distributions of the WFCs for all $j$ are
Gaussian (the central limit theorem). The amplitude $\delta$ can
only play the role as one variable of the field.

Moreover, even when the one-point distributions of 2$^j$ WFCs at
all $j$ are Gaussianized, the mass field could still be
non-Gaussian. For instance, suppose the one-point distribution of
the 2$^j$ WFCs, $\tilde{\epsilon}^M_{j,l}$, on scale $j$, is
Gaussian. If the WFCs on scale $j+1$ is given by
\begin{eqnarray}
\tilde{\epsilon}^M_{j+1;2l} & = & a\tilde{\epsilon}^M_{j,l}, \\
\nonumber \tilde{\epsilon}^M_{j+1;2l+1} & = &
b\tilde{\epsilon}^M_{j,l},
\end{eqnarray}
where $a$ and $b$ are arbitrary constant, the one-point
distribution of the 2$^{j+1}$ WFCs $\tilde{\epsilon}^M_{j+1,l}$ is
also Gaussian. However, Eq.(18) leads to a strong correlation
between $\tilde{\epsilon}^M_{j+1,l}$ and
$\tilde{\epsilon}^M_{j,l}$. This is an example of the scale-scale
correlation, i.e. the scale $j+1$ fluctuations are always
proportional to those on the scale $j$ at the same position.
Moreover, this correlation can not be eliminated by the
Gaussianization of $\tilde{\epsilon}^M_{j;l}$. The Gaussian
mapping changes all the WFCs {\it at a given position} (pixel) by
a same amplifying or reducing factor, and therefore, the local
relations Eq.(18) remains.

The scale-scale correlations only depend upon the statistical
behavior of the fluctuation distribution with respect to the index
$j$. Therefore, a Gaussian field requires the uncorrelation
between the distributions of WFCs with different $j$. This
uncorrelation corresponds to decorrelating the band average
Fourier modes, which will be discussed in detail in \S 4.

\subsection{Algorithms of scale-by-scale Gaussianization}

Based on the considerations in the last section, we may design an
algorithm which is capable of reducing the contamination of the
non-Gaussianity, and produce fields with less non-Gaussianity.

The new method is based on the scale-by-scale decomposition of
flux and mass field. From Eq.(6), we have
\begin{equation}
F = F^j + \sum_{j'=j}^{\infty} \sum_{l= 0}^{2^{j'} -1}
  \tilde{\epsilon}_{j',l} \psi_{j',l},
\end{equation}
and
\begin{equation}
F^j \equiv \bar{F} + \sum_{j'=0}^{j-1} \sum_{l= 0}^{2^{j'} -1}
  \tilde{\epsilon}_{j',l} \psi_{j',l}.
\end{equation}
$F^j$  is actually a smoothed $F$ by a filter on the scale $j$.
There is a recursion relation in $F^j$ given by
\begin{equation}
F^{j+1} = F^j +
  \sum_{l= 0}^{2^j -1}\tilde{\epsilon}_{j,l} \psi_{j,l}.
\end{equation}
Namely, flux $F^{j+1}$ can be reconstructed from flux $F^j$ and
2$^j$ WFCs $\tilde{\epsilon}_{j,l}$ at the scale $j$.  Similarly,
for a mass distribution, we have
\begin{equation}
\delta = \delta^j + \sum_{j'=j}^{\infty} \sum_{l= 0}^{2^{j'} -1}
  \tilde{\epsilon}^M_{j',l} \psi_{j',l},
\end{equation}
\begin{equation}
\delta^j \equiv \sum_{j'=0}^{j+1} \sum_{l= 0}^{2^{j'} -1}
  \tilde{\epsilon}^M_{j',l} \psi_{j',l},
\end{equation}
and
\begin{equation}
\delta^{j+1} = \delta^j +
  \sum_{l= 0}^{2^j -1}\tilde{\epsilon}^M_{j,l} \psi_{j,l}.
\end{equation}

Since the relations between $F$ and $\delta$ are local and
monotonic, the smoothed flux $F^{j+1}$ depends only on the
smoothed mass field $\delta^{j+1}$, and one can perform a local
and monotonic mapping between $F^{j+1}$ and $\delta^{j+1}$. Thus,
we can implement the reconstruction of the mass field
$\delta^{j+1}$ from $F^{j+1}$ by a scale-by-scale Gaussianization
algorithm (hereafter referred to as algorithm I):

1. Supposing the reconstruction down to the scale $j$ has been
done, i.e. the $\delta^j$ is known already;

2. Calculating the WFCs of the flux $F$ on the scale $j$;

3. Making the Gaussian mapping of the 2$^j$ WFCs
$\tilde{\epsilon}_{j,l}$, and assigning the Gaussianized result,
$\varepsilon_{j,l}$, to the 2$^j$ pixels according to the rank
order. The distribution of $\varepsilon_{j,l}$ is Gaussian with
zero mean and variance one.

4. Finding the 2$^j$ WFCs of mass field by
\begin{equation}
\tilde{\epsilon}^M_{j,l}=\nu \varepsilon_{j,l}.
\end{equation}
where the parameter $\nu$ is a normalization factor to be
determined. The one-point distribution of the WFCs of mass field
at the scale $j$, $\tilde{\epsilon}^M_{j,l}$, is then
Gaussianized.

5. Reconstructing the mass field $\delta^{j+1}$ on scale $j+1$ by
the recursion relation Eq.(24)

6. To determine the parameter $\nu$, we require that the DWT power
   spectrum of the flux $F^{j+1}$ simulated from $\delta^{j+1}$
   reproduces the observed flux $F^{j+1}$. We have then $\delta^{j+1}$.
   The reconstruction of mass field on the scale $j+1$ is done.

Repeating the steps 1 to 6, one can reconstruct the mass field on
scales from large to small until the scale of the resolution of
the flux $F$, or the scale on which the relation between $F$ and
$\delta$ is no longer local.

Figure 7 illustrates the transmitted flux, the initial density
field and the recovered density field by algorithm I. The
recovered 1D density field is in excellent agreement with the
original density field scale-by-scale. The non-Gaussianities of
the recovered fields by algorithm I are shown in Figs. 4 - 6. The
skewness and kurtosis spectra exhibit almost nothing but
Gaussianity. The scale-scale correlation is also significantly
reduced.

The Gaussianization algorithm I is conceptually clear. However, it
needs to determine the normalization factor $\nu$ at each scale.
Therefore, it is rather cumbersome to do the numerical calculation.
Moreover, there is still somewhat residual scale-scale correlation
in the recovered mass field. In fact, algorithm I does ensure the
Gaussian PDF of $\tilde{\epsilon}^M_{j,l}$, but it is unable to
remove all the correlation between different modes, just as the
simple example [Eq.(18)] demonstrated in \S 3.2.

To avoid the multiple normalizations and keep the virtues of
scale-by-scale Gaussianization, we design an alternative algorithm
as follows (hereafter referred to as algorithm II):

1. Using the conventional Gaussianization (\S 3.1) to reconstruct
the mass field, i.e. to perform Gaussian mapping of the density
contract $\delta$ and normalize the mass field by requiring that
the evolved simulations reproduce the power spectrum of the
observed flux.

2. Calculating the WFCs $\tilde{\epsilon}_{j,l}$ of the recovered
mass field $\delta^M$ on each scale $j$.

3. Similar to the step 3 of algorithm I, making the
Gaussianization of
   $\tilde{\epsilon}_{j,l}$ for each scale $j$ to produce
   unnormalized WFCs $\varepsilon^M_{j,l}$.

4. Normalizing the WFCs $\varepsilon^M_{j,l}$ on scale $j$ by
   requiring that the variance of $\varepsilon^M_{j,l}$, i.e., the 2nd
   cumulant moment $I_{j}^2$ [Eq.(7)], is the same as those for the WFCs
   $\tilde{\epsilon}_{j,l}$.

5. For each scale $j$, randomizing the spatial sequence of the
Gaussianized WFCs $\varepsilon^M_{j,l}$, i.e., making a random
permutation among the index $l$.

6. Using these WFCs $\varepsilon^M_{j,l}$, one can reconstruct the
mass density field by Eq.(24) till the scale given by the
resolution of the flux.

Algorithm II is still scale-by-scale in nature. However, the
normalization is one only once for the recovered $\delta$. The
step 4 ensure that the normalization is unchanged after the step 3,
which eliminates the non-Gaussianities of the skewness and kurtosis
spectra. Step 5 is for eliminating the residual scale-scale
correlations by a randomization of the spatial index $l$ of
$\tilde{\epsilon}_{j,l}$. Namely, it changes only the position of
$\tilde{\epsilon}_{j,l}$, but not the values. Therefore, it
is similar to a randomization of phases of the Fourier modes, and
will not change the normalization of the amplitude and
power spectrum of the fields.

Figs. 4 - 6 show that the Gaussianized field by the algorithm II
contains almost none of the non-Gaussian features considered. However,
it should be pointed out that the field given by algorithm II is no
longer a point-to-point reconstruction due to the randomization of
$l$. Namely, the recovered field will not be point-to-point the same
as the field shown in Fig. 7. Nonetheless, since the purpose of the
Gaussianization is to recover the power spectrum of the primordial
density fluctuations, algorithm II is a valuable approach. As will
be shown in next section, the algorithm II gives more unbiased
estimation of power spectrum by the standard FFT technique.

In order to illustrate the effect of the peculiar velocities on the
Gaussianization, each of Figs. 4 - 6 contains two panels: one employed the
simulation samples including the effects of peculiar velocities, and the other
did not. All the figures show that for the algorithm I, the effect of peculiar
velocities is significant only on small scales $j>9$, or $k>10$ h Mpc$^{-1}$;
while for algorithm II, the effect of peculiar velocities appears on smaller
scales. Therefore, our proposed scale-by-scale Gaussianization methods would
not be affected by the peculiar velocities as least up to the scale j=9.

\section{Recovery of mass power spectrum from the transmitted QSO flux}

\subsection{The power spectrum in different representations}

As a preparation for measuring the non-Gaussian effects on power
spectrum recovery, we first discuss the representation of power
spectrum. Principally, a random field can be described by any
complete orthonormal basis (representation). Although the default
usage of power spectrum is defined on the Fourier basis, one can
define the power spectrum with respect to different
representation. This is due to the fact that the Parseval's
theorem holds for any complete and orthonormal basis
decomposition.

In the Fourier representation, the power spectrum of a 1-D density
field $\delta(x)$ is given by
\begin{equation}
P(n)= |\hat{\delta}_n|^2.
\end{equation}
where $\hat{\delta}_n$ is the Fourier transform of $\delta(x)$.
$|\hat{\delta}_n|^2$ measures the power of mode $n$ because of
Parseval's theorem
\begin{equation}
\frac{1}{L} \int_0^L \delta^2(x) dx =
   \sum_{n= -\infty}^{\infty} |\hat{\delta}_n|^2.
\end{equation}

Similarly, we have the Parseval's theorem for the DWT transform
given by (Fang \& Thews 1998, Pando \& Fang, 1998b)
\begin{equation}
\frac{1}{L} \int_{0}^{L} \delta^2(x) dx = \sum_{j=0}^{\infty}
\frac{1}{L} \sum_{l = 0}^{2^j-1} \tilde{\epsilon}^2_{j,l}.
\end{equation}
(For simplicity, we ignore the superscript $M$ on
$\tilde{\epsilon}_{j,l}$). Therefore, the term
$\tilde{\epsilon}^2_{j,l}$ describes the power of mode $(j,l)$,
and the total power on the scale $j$ is
\begin{equation}
P_j =\frac{1}{L}\sum_{l=0}^{2^j-1} |\tilde{\epsilon}_{j,l}|^2,
\end{equation}
which defines the DWT power spectra $P_j$.

Generally, the second order correlation functions of
$\hat{\delta}_n$ or $\tilde{\epsilon}_{j,l}$ can be converted from
each other by
\begin{equation}
\langle\tilde{\epsilon}_{j,l}\tilde{\epsilon}_{j',l'}\rangle =
\sum_{n, n'=-\infty}^{+\infty}
\langle\hat{\delta}_n\hat{\delta}^*_{n'}\rangle
\hat{\psi}_{j',l'}(n')\hat{\psi}^*_{j,l}(n)
\end{equation}
\begin{equation}
\langle\hat{\delta}_n\hat{\delta}^*_{n'}\rangle = \sum_{j,
j'=0}^{+\infty}\sum_{l=0}^{2^j-1}\sum_{l'=0}^{2^{j'}-1}
\langle\tilde{\epsilon}_{j,l}\tilde{\epsilon}_{j',l'}\rangle
\hat{\psi}_{j,l}(n)\hat{\psi}^*_{j',l'}(n')
\end{equation}
where $\hat{\psi}_{j,l}(n)$ is the Fourier transform of
$\psi_{j,l}(x)$.
For an homogeneous random field,
$\langle\hat{\delta}_n\hat{\delta}^*_{n'}\rangle=
\langle|\hat{\delta}_n|^2\rangle\delta_{n, n'}$, we have then
\begin{equation}
\langle\tilde{\epsilon}^2_{j,l}\rangle =
\sum_{n=-\infty}^{+\infty}
\langle|\hat{\delta}_n|^2\rangle|\hat{\psi}_{j,l}(n)|^2
\end{equation}
or
\begin{equation}
P_j= \sum_{n=-\infty}^{+\infty} P(n) |\hat{\psi}(n/2^j)|^2
\end{equation}
where $\hat{\psi}(n/2^j)$ is the Fourier transform of the
generating wavelet $\psi(x)$ (Pando \& Fang 1998b). In Eq. (33)
the function $|\hat{\psi}(n/2^j)|^2$ plays the role of window
function in the wavenumber $n$ space. The function $\hat{\psi}(n)$
is localized in $n$-space. For the Daubechies 4 wavelet,
$|\hat{\psi}(n)|$ is peaked at $n=\pm n_p$ with the width of
$\Delta n_p$. Therefore, the DWT spectrum $P_j$ gives an estimator
of the ``band averaged" Fourier power spectrum within the band
centered at
\begin{equation}
\log n  = (\log 2)j + \log n_p,
\end{equation}
with the band width,
\begin{equation}
\Delta \log n=\Delta n_p/n_p.
\end{equation}

As the mean of the WFCs $\tilde{\epsilon}_{j,l}$ over $l$ is zero,
so the second cumulant moment $I_j^2$ is related to DWT spectrum
$P_j$ by $I_{j}^2=(L/2^j)P_j$, we will use variance $I_j^2$ as the
estimator of DWT power spectrum instead of $P_{j}$.

For a Gaussian field, the statistical behavior are completely
determined by the second order statistics of the Gaussian
variables $\hat{\delta}_n$ or $\tilde{\epsilon}_{j,l}$.
Theoretically, the power spectrum estimators  $P(n)$ and $P_j$
present the equivalent description. However, as will be shown
below, once non-Gaussianity appears, these estimators will no
longer be equivalent.

\subsection{Effect of non-Gaussianity on the recovery of mass
   power spectrum}

Using the 100 realizations of the mass density fields recovered
by the conventional algorithm, and algorithm I and II of the
Gaussianization, we calculated the power spectra by the standard
FFT technique. To reveal the effect of non-Gaussianity on the power
spectrum estimation, we do not include the effects of instrumental
noise and continuum fitting in the synthetic spectra. The dominant
sources of error in estimation of power spectrum would be the
cosmic variance and the non-Gaussian effects.

Figure. 8 compares the power spectra obtained by different
Gaussianization methods. The 1-D linear power spectrum of Eq.(3)
is also shown by solid line. These power spectra are normalized to
the present. In general, the recovered power spectrum can match
the shape of the linear theory over a wide range of wavelength,
especially on larger wavelength. Yet, the recovered spectra show
somewhat systematic departure from the initial mass power spectrum
with the increase of wavenumbers. For the conventional
Gaussianization, the recovered power spectrum falls below the
initial power spectrum on scales of $j\ge 8$ or $k\sim 1.5$
h$^{-1}$ Mpc. The power spectrum recovered by the algorithm I is
better than the conventional Gaussianization, and the recovery by
algorithm II gives the best one, which is almost the same as the
initial power spectrum on all scales.

Comparing Fig. 8 with Figs. 4 - 6, we can see that the scales on
which the depression of the recovered power spectrum appears is
always the same as the scale on which the scale-scale correlations
become significant. Moreover, the less the scale-scale correlation
(Fig. 6), the less the depression. This indicates that the
recovered spectrum is substantially affected by the
non-Gaussianities, especially, the scale-scale correlations.
Actually, this effect has already been recognized by Meiksin \&
White (1998) in analyzing N-body simulation samples. Namely, the
goodness of a power spectrum estimation is significantly dependent
on the correlation between the Fourier power spectra averaged at
different scale bands.

Back to the definition of scale-scale correlation Eq. (15), and
recall that the average over an ensemble is equivalent to the
spatial average taken over one realization, Eq.(15) can be rewritten as
\begin{equation}
C_j^{2,2}=
\frac{\langle\hat{P}_{j}\hat{P}_{j+1}\rangle}
{\langle\hat{P}_{j}\rangle\langle\hat{P}_{j+1}\rangle}.
\end{equation}
Hence, the scale-scale correlation is actually a measure of the
correlation between the Fourier power spectra averaged at different
scale bands. This can also be seen from Eq.(31) that the Fourier
power spectrum around $n$ depends on the fluctuations on different
$j$, and therefore, their non-Gaussian correlations.

Because the algorithm II is most effective for eliminating the
scale-scale correlations, the resulting power spectrum shows the best
recovery of the linear model.

\subsection{Suppression of non-Gaussian correlations by representation}

In the DWT representation, the power spectrum (29) does not depend
on modes at the scales different from $j$, and therefore the
scale-scale correlation will not affect the estimation of $P_j$.
One can expect that the DWT spectrum estimator, $P_j$, will give a
better recovery of the initial power spectrum.

Fig. 9 displays the DWT power spectrum $P_j$ for mass fields given
by the different Gaussianization methods. The DWT power spectrum in
the linear CDM model is also shown by solid line, which is calculated
from the Fourier linear power spectrum by Eq. (33) in the continuous
limit of $n$. This figure indicates that even for the mass field
recovered from conventional Gaussianization, the DWT power spectrum
is in good agreement with the initial DWT mass power spectrum up to
the scale $j=9$. This is already much better than its counterpart in
the Fourier representation, for which the power spectrum shows
significant difference from the linear spectrum on scale $j \sim 8$.
For the algorithm I and II, the DWT power spectrum also gives the
good results. In addition, the errors due to the cosmic variance
and normalization in the DWT spectrum are manifestly smaller than
that of Fourier spectrum.

The DWT power spectrum $P_j$ (29) is given by the summation of
$|\tilde{\epsilon}_{j,l}|^2$ over $l$ at a given scale. Therefore,
the non-Gaussian effect on estimation of $P_j$ mainly arises from
the correlation between the WFCs $\tilde{\epsilon}_{j,l}^2$ at
{\it different} $l$, which can be measured by
\begin{equation}
Q^{2,2}_{j,\Delta l} = \frac{2^{j}\sum_{l=0}^{2^{j}-1}
\tilde{\epsilon}^2_{j;l} \; \tilde{\epsilon}^2_{j;l+\Delta l}}
{\sum \tilde{\epsilon}^2_{j,l} \sum \tilde{\epsilon}^2_{j;l+\Delta
l}}.
\end{equation}
$Q^{2,2}_{j,\Delta l}$ gives the correlation between the density
fluctuations on the same scale $j$ at different places $l$ and
$l+\Delta l$.

Fig. 10 displays the correlations $Q^{2,2}_{j,\Delta l}$ with
$\Delta l=1$. It shows that this non-Gaussianity can be ignored
till $j=10$. On the other hand, the scale-scale correlation
$C^{2,2}_j$ had been significant on $j=8$ (Fig. 3). In result,
the Fourier power spectrum is contaminated by the non-Gaussianity on
$j \geq 8$, while the DWT power spectrum is less biased till $j =9$.

In a word, the non-Gaussian correlations are effectively
suppressed in the DWT representation. The DWT spectrum estimator
gives a better recovery of the initial power spectrum.

\section{Conclusions}

In the cosmological reconstruction of initial Gaussian mass power
spectrum, a serious obstacle is the non-Gaussianity of the evolved
field. The quality of the recovery of the power spectrum is affected
by the non-Gaussian correlations. The precision to which the mass
power spectrum could be measured relies on how to treat the
non-Gaussianity of the evolved mass field.

In the quasi-nonlinear regime of cosmic gravitational clustering
(like that traced by the Ly$\alpha$ forests), the dynamical
evolution is characterized by the power transfer from large scale
perturbations to small ones (Suto \& Sasaki 1991). This is the
mode-mode coupling which produces the scale-scale correlations.
Using perturbation theory in the DWT representation, one can
further show that the mode-mode coupling at the same position
(local coupling) is much stronger than coupling between modes at
different positions (non-local) (Pando, Feng \& Fang 1999). On the
other hand, the power spectrum in the quasi-nonlinear regime does
not significantly differ from the linear regime. Therefore, the
algorithm for recovering the initial mass power spectrum from the
Ly$\alpha$ forests should be designed to eliminate the local
scale-scale correlations of the evolved mass field.

Using simulations in semi-analytical model of the Ly$\alpha$ forests,
we show that the conventional algorithm of the Gaussianization is
not enough to recover a Gaussian field. The local scale-scale
correlations of the Ly$\alpha$ forests are still retained in the
Gaussianized mass field. Based on the DWT scale-space decomposition,
we proposed two algorithms of the Gaussianization, which are
effective to eliminate the non-Gaussian features.

We showed that representation selection is important for the
recovery of the power spectrum. A representation, which can
effectively suppress the contamination of local scale-scale
correlations, would be good for extracting the initial linear
spectrum. We compared the Fourier and DWT representations for the
estimation of power spectrum. We demonstrated that, at least in
the quasi-nonlinear regime, the DWT power spectrum estimator is
better, because it can avoid the major contamination, the local
scale-scale correlations. We also showed that the peculiar
velocities of gas will not affect on the DWT power spectrum
recover up to, at least, the scale $j=9$.

\acknowledgments

We thank Dr. D. Tytler for kindly providing the data of the Keck
spectrum HS1700+64. We also thank Drs. Wei Zheng, Hongguang Bi and
Wolung Lee for useful discussion. LLF acknowledges support from
the National Science Foundation of China(NSFC) and World
Laboratory scholarship. This project was done during LLF's
visiting at Department of Physics, University of Arizona. This
work was supported in part by LWL foundation.

\newpage

\begin{figure}
\figurenum{1} \epsscale{0.8} \plotone{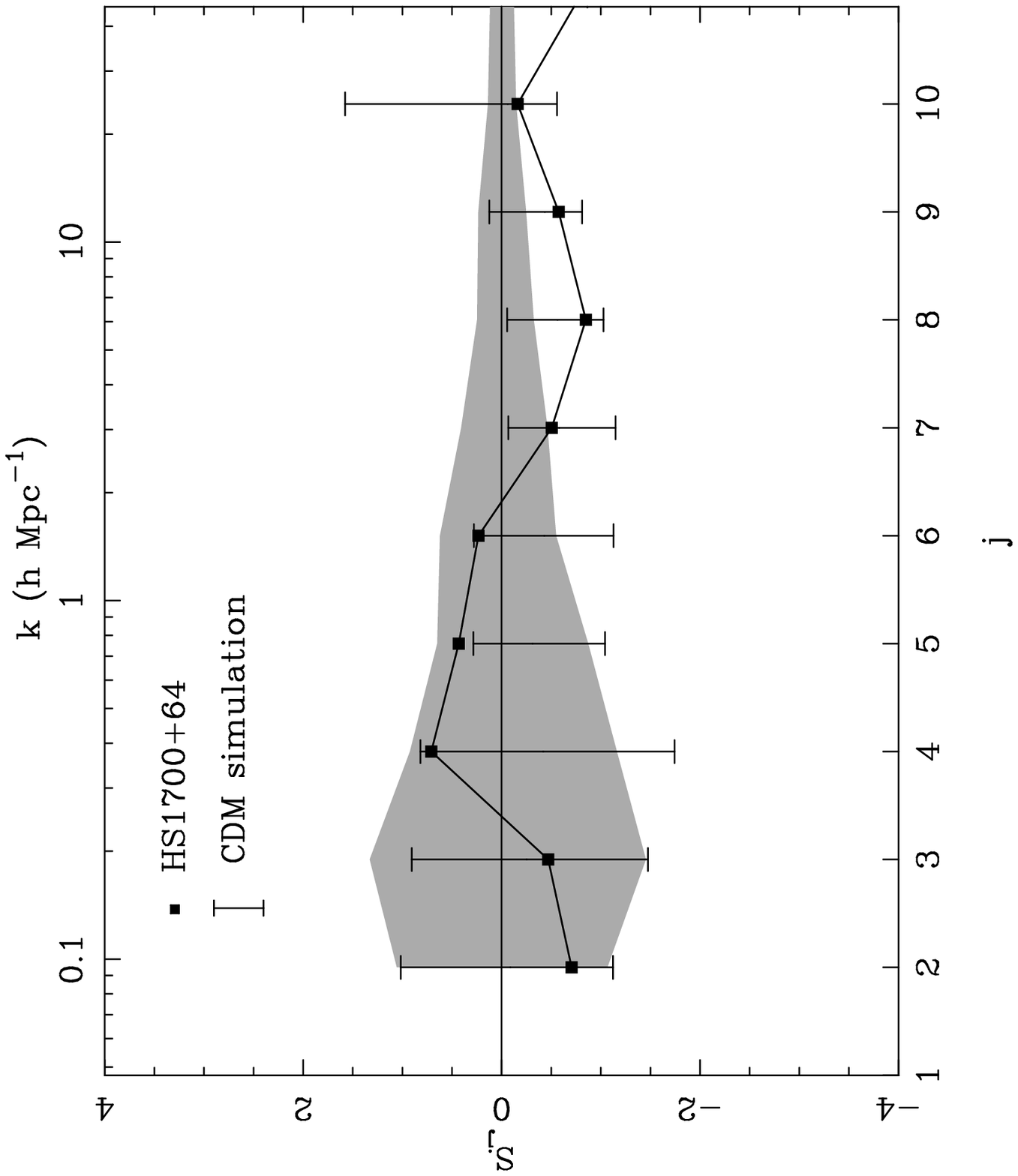} \caption{The
skewness spectrum of the Ly$\alpha$ forests. The 95\% confidence
ranges of the skewness spectrum of the simulated samples and
Gaussian noise are shown by the bars and gray band, respectively.
The skewness spectrum of the Keck data of HS1700+64 is shown by
squares and solid line. The physical scale related to $j$ is
$189.8\times 2^{-j}$ h$^{-1}$ Mpc in the CDM model.}
\end{figure}

\begin{figure}
\figurenum{2} \epsscale{0.8} \plotone{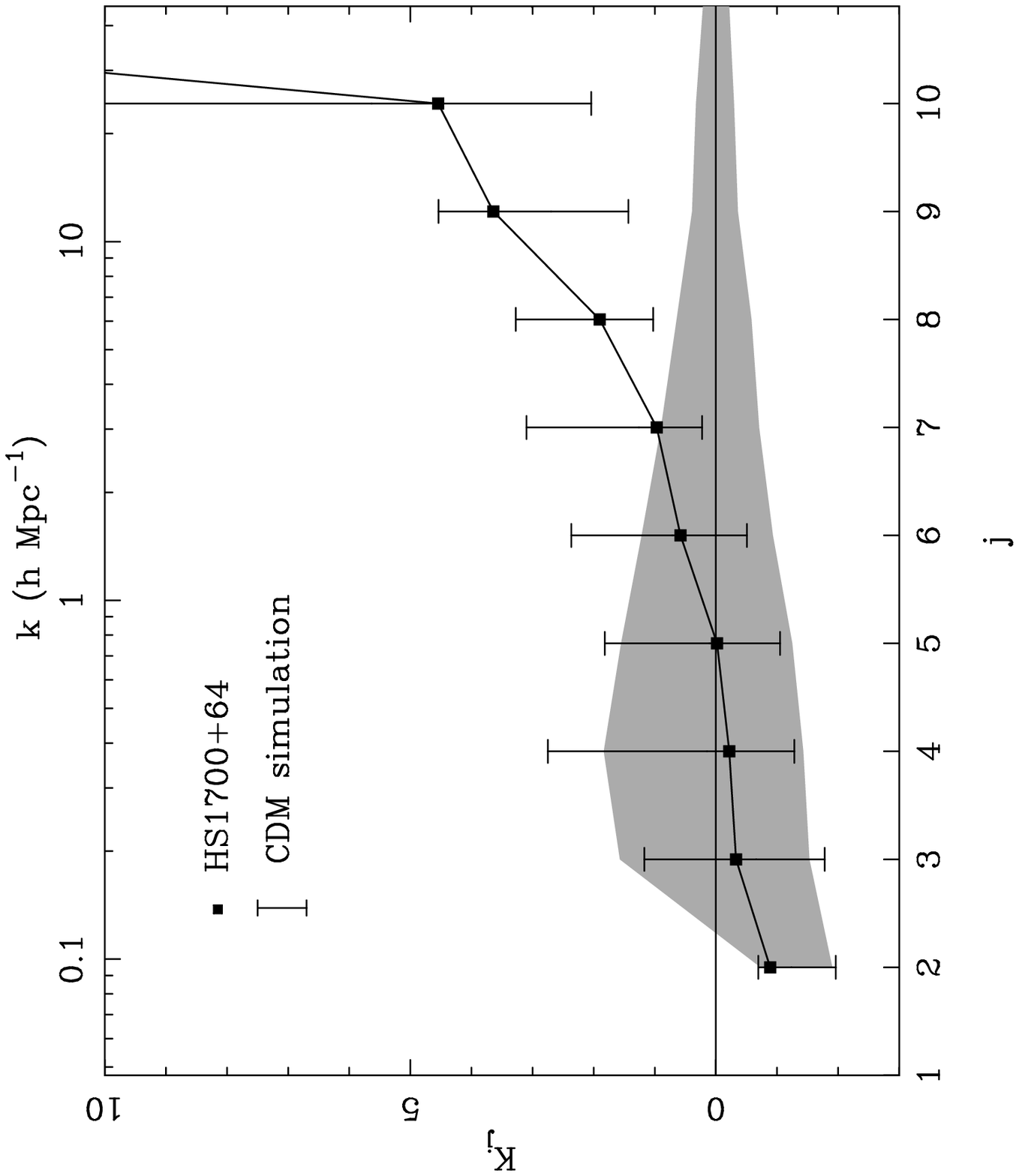} \caption{The
kurtosis spectrum of the Ly$\alpha$ forests. The 95\% confidence
ranges of the kurtosis spectrum of the simulated samples and
Gaussian noise are shown by the bars and grey band, respectively.
The kurtosis spectrum of the Keck data of HS1700+64 is shown by
squares and solid line. The physical scale related to $j$ is
$189.8\times 2^{-j}$ h$^{-1}$ Mpc in the CDM model.}
\end{figure}

\begin{figure}
\figurenum{3} \epsscale{0.8} \plotone{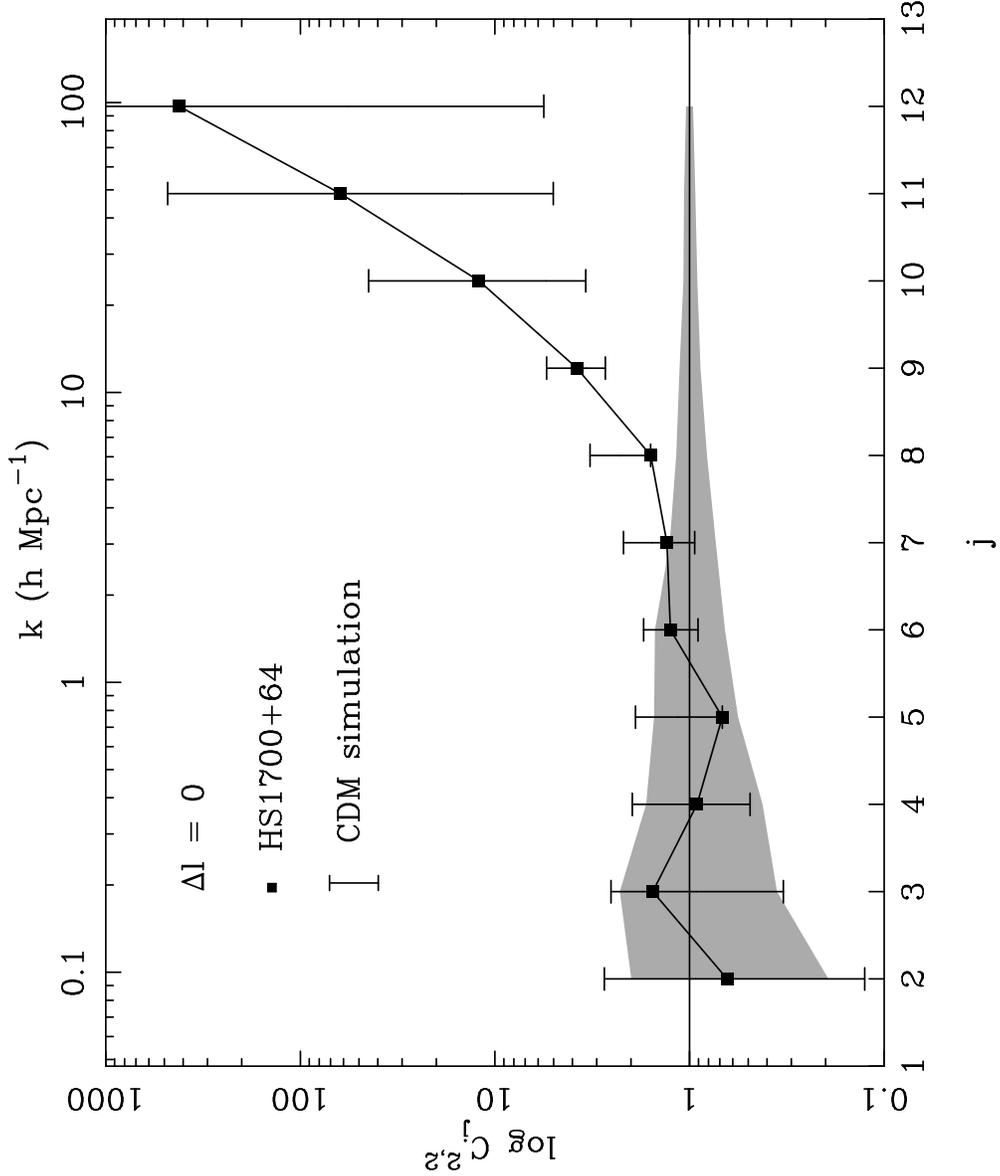} \caption{The
scale-scale correlation of the Ly$\alpha$ forests. The 95\%
confidence ranges of the scale-scale correlation, $C_{j}^{2,2}$,
for the simulated samples and Gaussian noise are shown by the bars
and grey band, respectively. The scale-scale correlation of the
Keck data of HS1700+64 is shown by squares and solid line. The
physical scale related to $j$ is $189.8\times 2^{-j}$ h$^{-1}$ Mpc
in CDM model.}
\end{figure}

\begin{figure}
\figurenum{4} \epsscale{0.8} \plotone{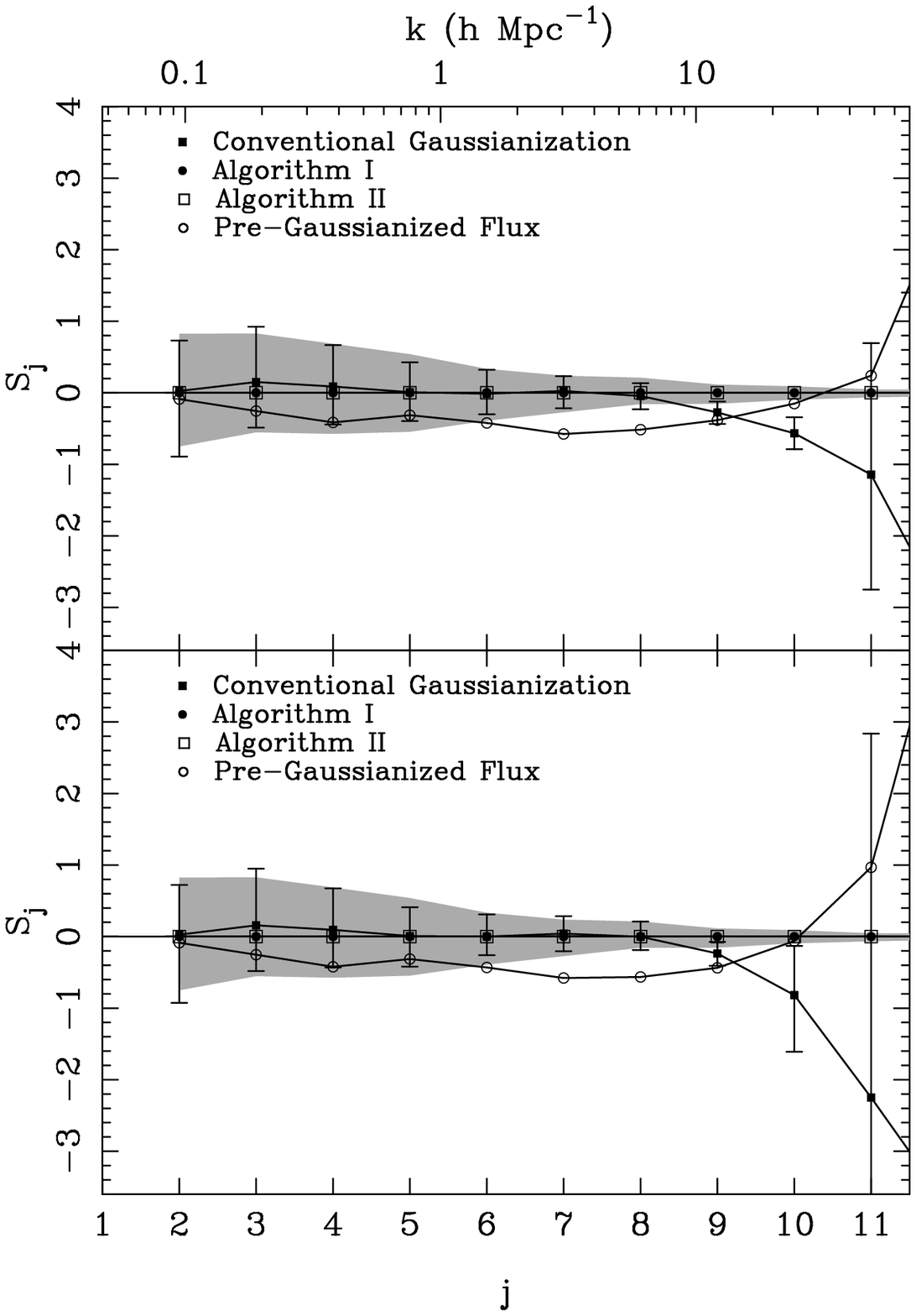} \caption{The
skewness spectrum of mass field recovered by the three
Gaussianization methods. The upper panel is for transmitted flux
samples free from the effect of peculiar velocity, and the lower
panel is for samples affected by the peculiar velocity. The 95\%
confidence ranges of the skewness spectrum are shown by the
symbols indicated in figure. The gray band is the 95\% confidence
ranges for the Gaussian noise samples.}
\end{figure}

\begin{figure}
\figurenum{5} \epsscale{0.8} \plotone{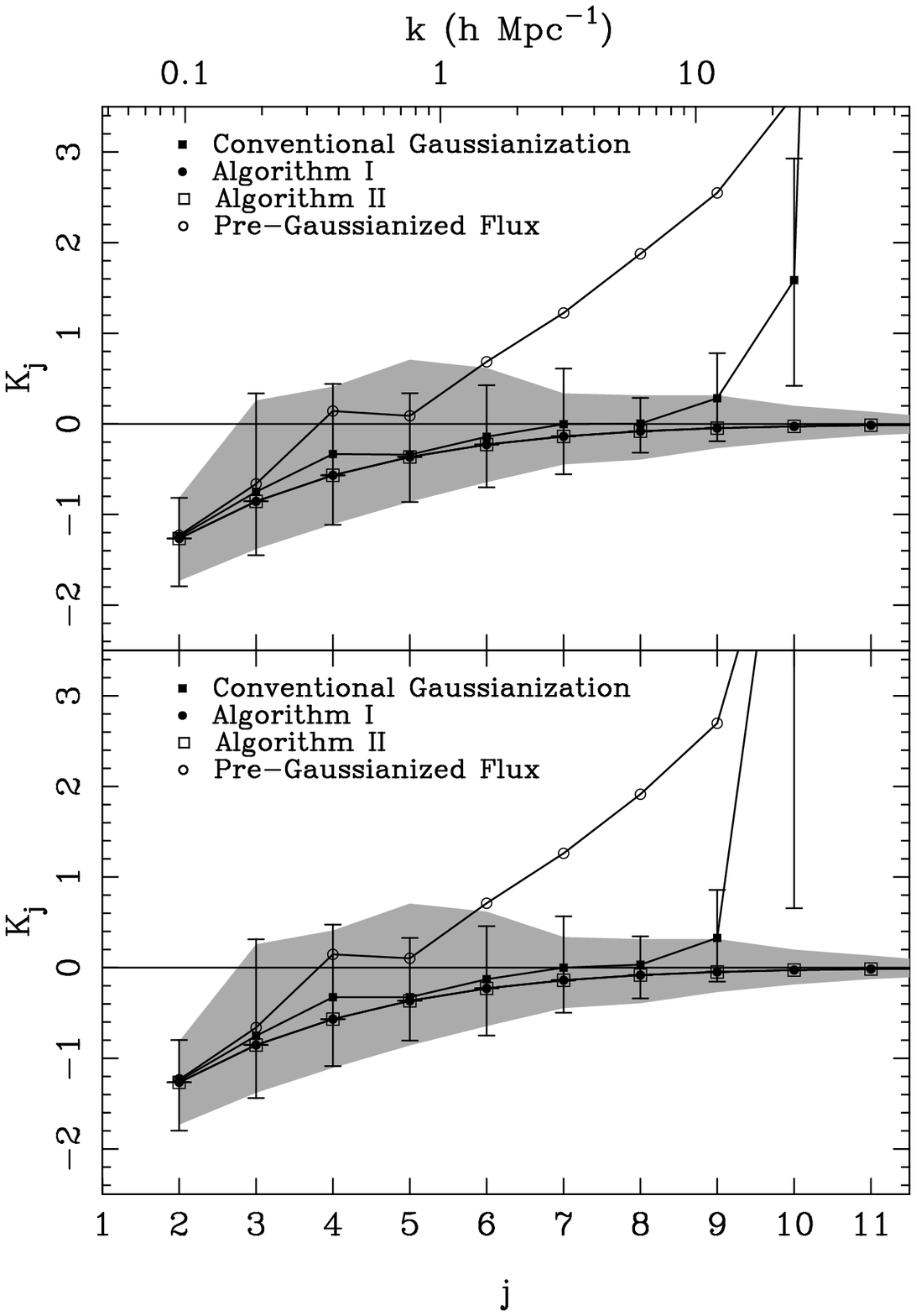} \caption{The
kurtosis spectrum of mass field recovered by the three
Gaussianization methods. The upper panel is for transmitted flux
samples free from the effect of peculiar velocity, and the lower
panel is for samples affected by the peculiar velocity. The 95\%
confidence ranges of the skewness spectrum are shown by the
symbols indicated in figure. The gray band is the 95\% confidence
ranges for the Gaussian noise samples.}
\end{figure}

\begin{figure}
\figurenum{6} \epsscale{0.8} \plotone{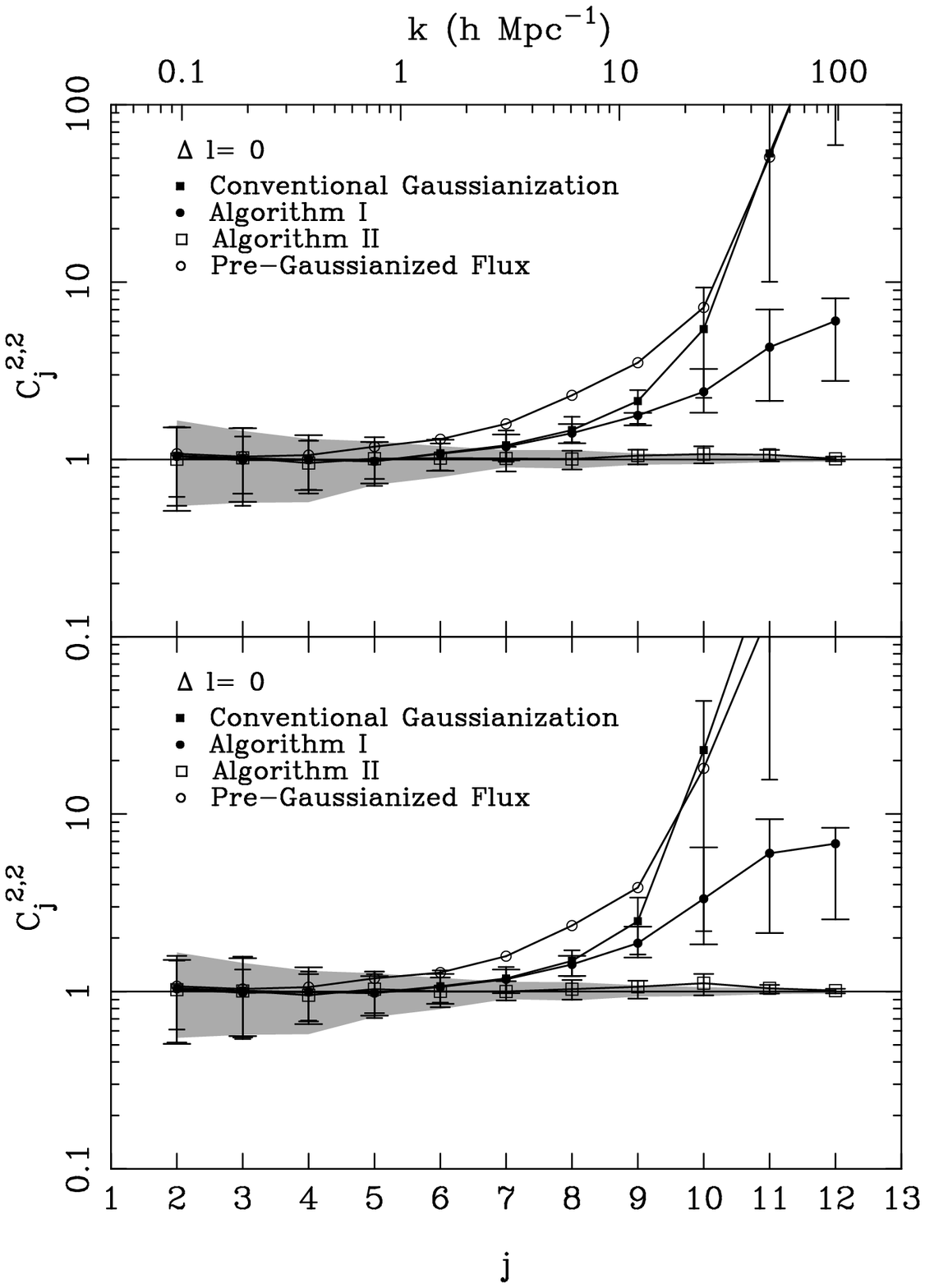} \caption{The
scale-scale correlation of mass field recovered by the three
Gaussianization methods. The upper panel is for transmitted flux
samples free from the effect of peculiar velocity, and the lower
panel is for samples affected by the peculiar velocity. The 95\%
confidence ranges of the skewness spectrum are shown by the
symbols indicated in figure. The gray band is the 95\% confidence
ranges for the Gaussian noise samples.}
\end{figure}

\begin{figure}
\figurenum{7} \epsscale{0.64} \plotone{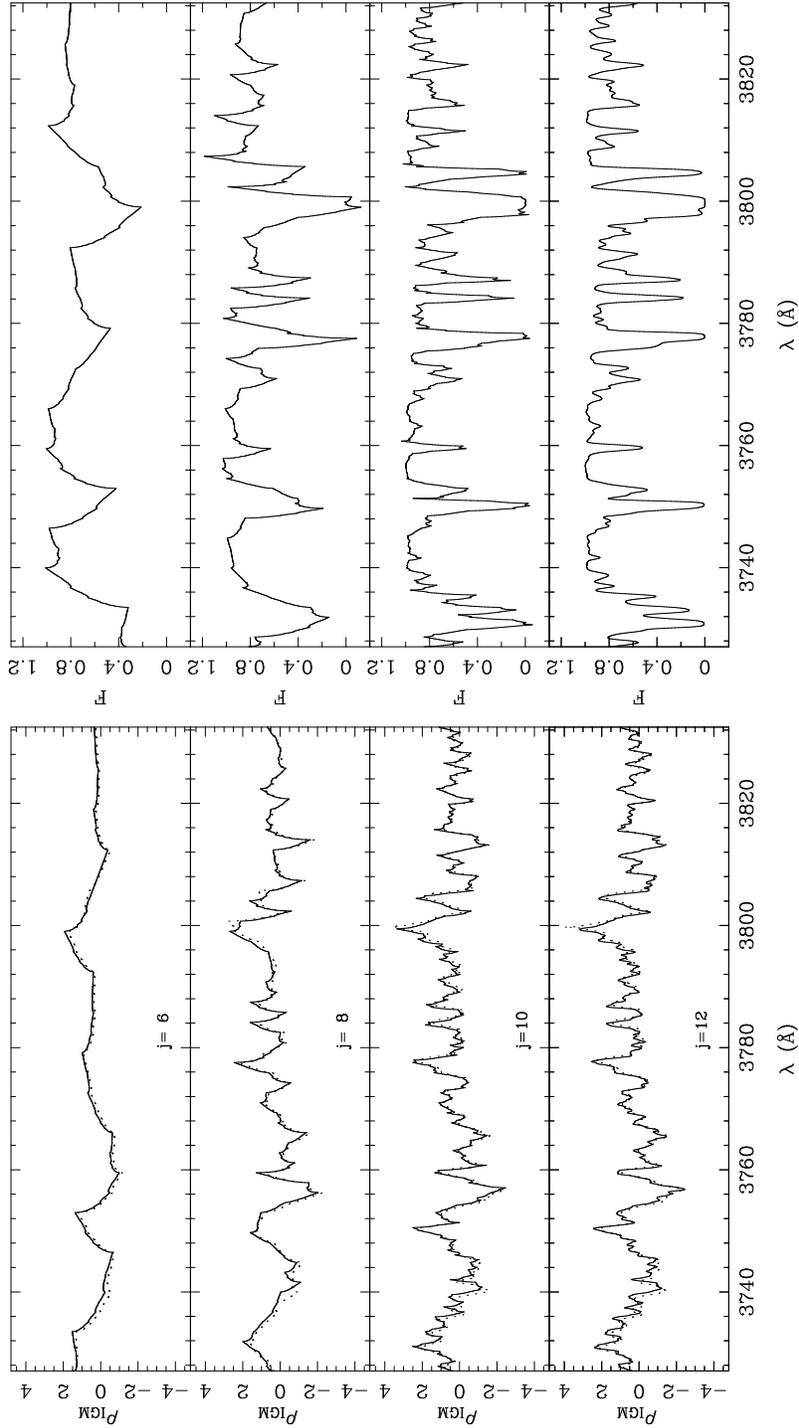} \caption{An
example of the density field (left panel) and flux (right panel)
reconstructed by the scale-by-scale Gaussianization (left panel).
The solid lines are for the reconstructed density field and flux.
The dot lines are for the original density field and flux.
Actually, the solid line and dot line are coincident in most
places.}
\end{figure}

\begin{figure}
\figurenum{8} \epsscale{0.75} \plotone{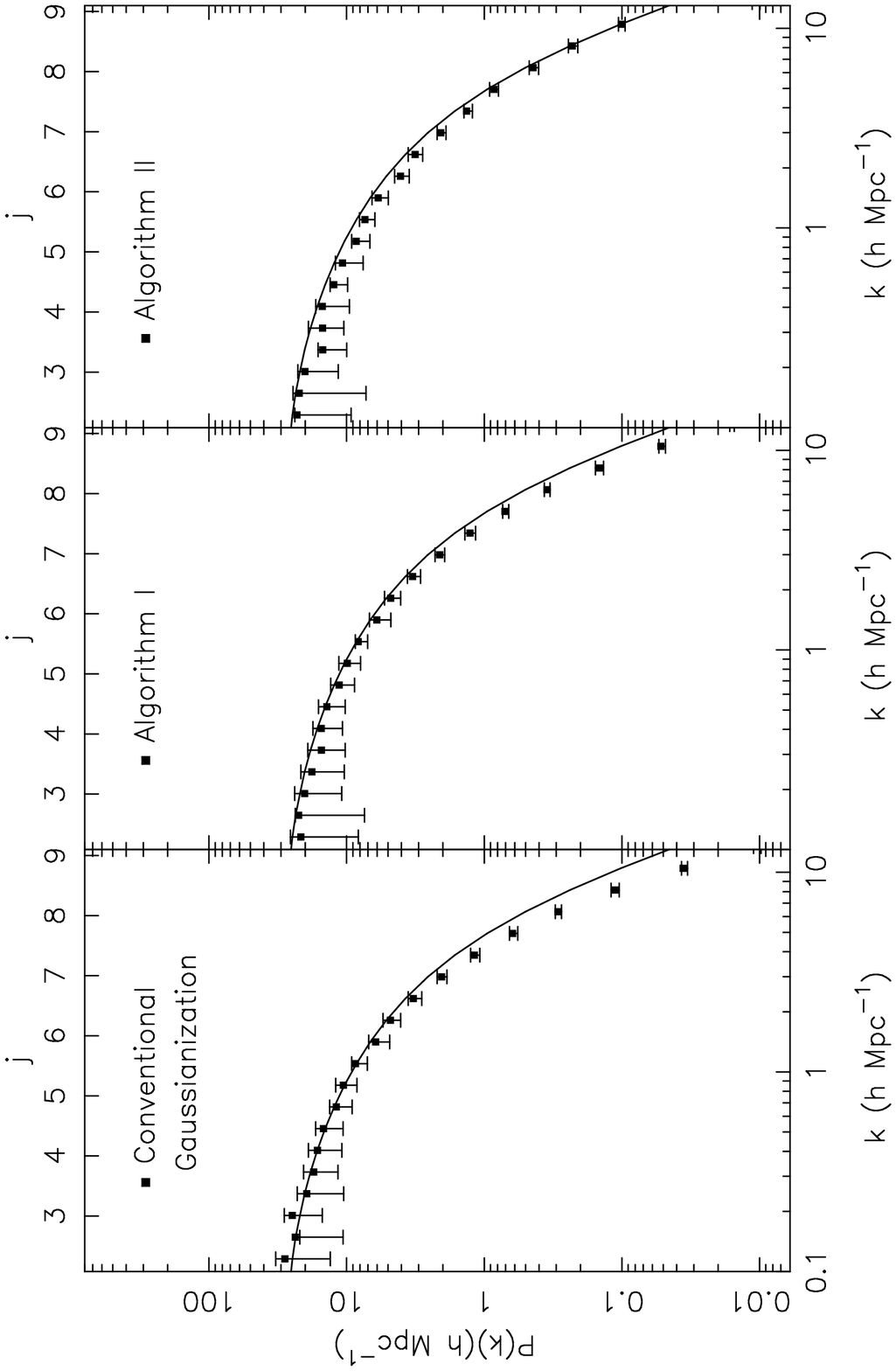} \caption{The
Fourier power spectrum of the mass fields recovered by the
conventional Gaussianization (left panel), algorithm I (central
panel) and algorithm II(right panel). The error bars are given by
$1\sigma$ deviation calculated for the 100 realizations. The
Fourier power spectrum of the linear CDM mode smoothed by Jeans
filter [Eq.(3)] is shown by solid line.}
\end{figure}

\begin{figure}
\figurenum{9} \epsscale{0.75} \plotone{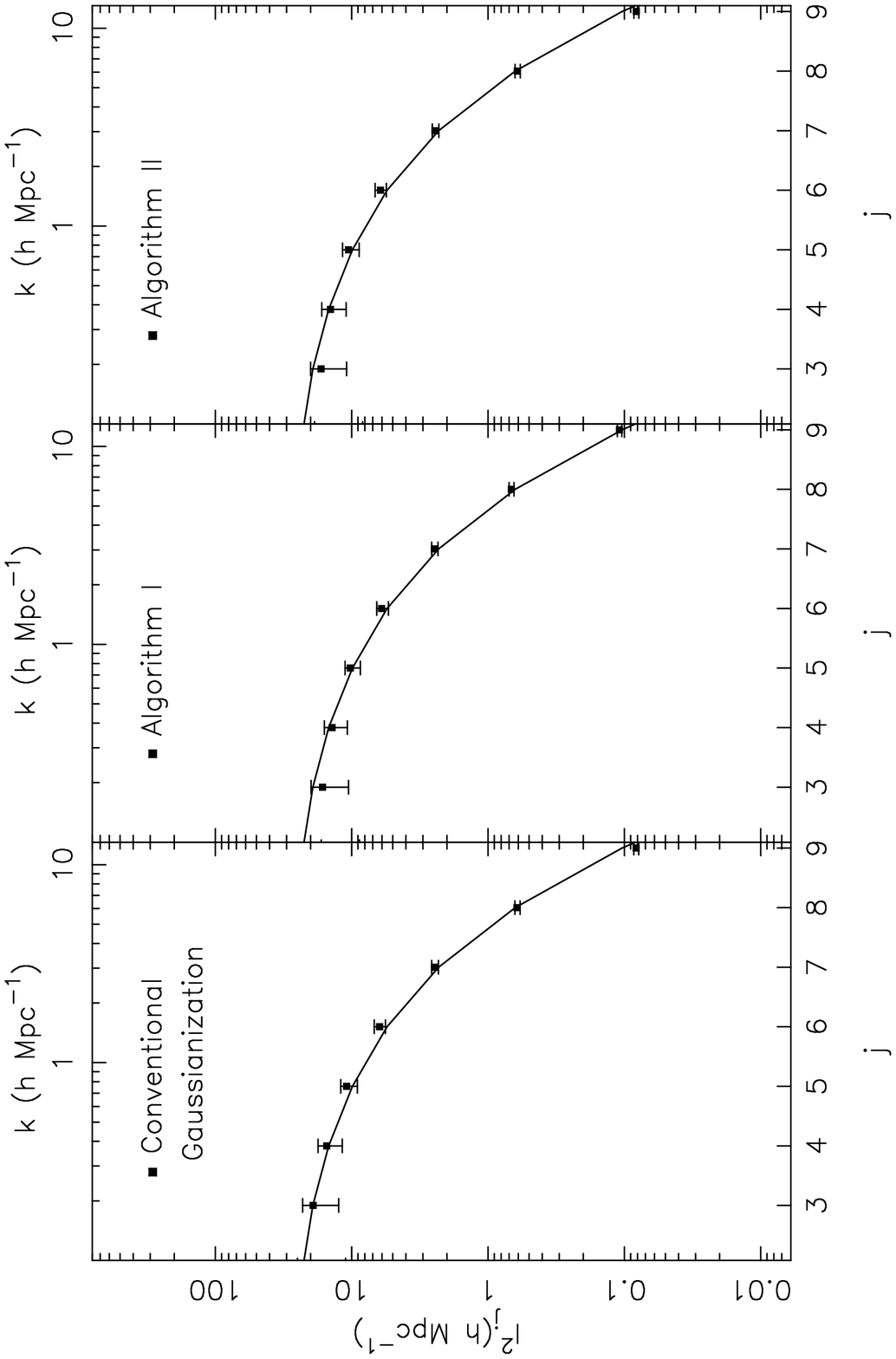} \caption{The DWT
power spectra of the mass fields recovered by the conventional
Gaussianization (left panel), algorithm I (central panel) and
algorithm II(right panel). The error bars are given by $1\sigma$
deviation calculated for the 100 realizations. The DWT power
spectrum of the linear CDM mode smoothed by Jeans filter [Eq.(3)]
is shown by solid line.}
\end{figure}

\begin{figure}
\figurenum{10} \epsscale{0.8} \plotone{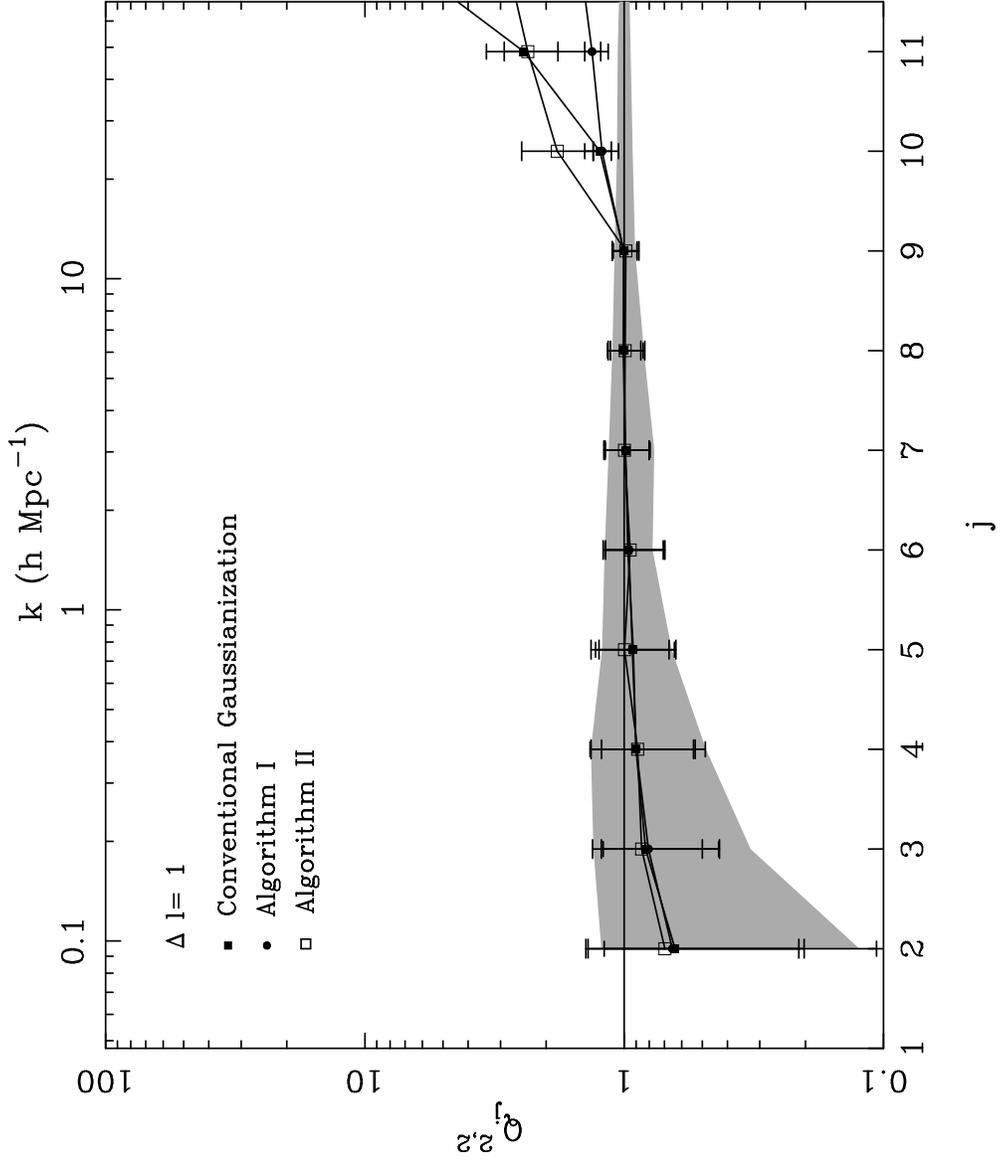} \caption{The
non-local correlations $Q_{j,\Delta l}^{2,2}$ with $\Delta l =1$
of the recovered density field by the three Gaussianization
methods. The symbols displayed in the figure have the same
meanings as Fig. 6. }
\end{figure}

\end{document}